\documentstyle[12pt]{article}

\def\prd#1{{\em Phys.~Rev.}~{\bf D#1}\ }
\def\prl#1{{\em Phys.~Rev.~Lett.}~{\bf #1}\ }
\def\plett#1{{\em Phys.~Lett.}~{\bf #1B}\ }
\def\np#1{{\em Nucl.~Phys.}~{\bf B#1}\ }
\def\deg{\ifmmode{^{\circ}}\else ${^{\circ}}$\fi}

\def\bi{\begin{itemize}}
\def\ei{\end{itemize}}
\def\ed{\end{document}}
\def\be{\begin{equation}}
\def\ee{\end{equation}}
\def\beq{\begin{eqnarray}}
\def\eeq{\end{eqnarray}}
\def\vev#1{\left<{#1}\right>}
\def\bm#1{\ifmmode{\mbox{\boldmath $#1$}}\else {\boldmath $#1$}\fi}

\def\mp{M_{\rm P}}
\def\mgut{M_{\rm GUT}}
\def\st{{\bf 16} }
\def\at{{\tilde\alpha}}
\def\sg{_{\rm G}}
\def\atg{\at\sg}
\def\tfrac#1#2{{\textstyle\frac{#1}{#2}}}

\def\gev{\ \mbox{GeV}}

\def\mev{\ \mbox{MeV}}
\def\pri{^{\, \prime}}

\def\eps{\epsilon}

\def\eb{\end{thebibliography}}

\def\mueg{\ifmmode{\mu\rightarrow e\gamma} \else $\mu\rightarrow e\gamma$\fi}
\def\taumg{\ifmmode{\tau\rightarrow \mu\gamma}\else
$\tau\rightarrow \mu\gamma$\fi}
\def\taueg{\ifmmode {\tau\rightarrow e\gamma}\else
$\tau\rightarrow e\gamma$\fi}
\def\mg{M\sg}
\def\rc{{\cal R}}
\def\tp{t_{\rm P}}
\def\cw{c_W}
\def\cwt{\cw^2}
\def\sw{s_W}
\begin{document}
\begin{titlepage}
\begin{flushright}  {\sl NUB-3128/95-Th}\\   {\sl Oct 1995}\\   hepph/9510303
\end{flushright}
\vskip 0.5in
\begin{center}   {\Large\bf Lepton Flavor Violation in SUSY-SO(10)
with Predictive Yukawa Texture}\\[.5in]   {Mario E. G\'{o}mez and Haim
Goldberg}\\[.1in]   {\sl Department of Physics}\\
{\sl Northeastern University}\\
{\sl Boston, MA 02115}
\end{center}
\vskip 0.4in
\begin{abstract}
We analyze the scalar lepton mass matrices in a supersymmetric SO(10) grand
unified model with soft SUSY breaking terms generated at Planck scale and a
Georgi-Jarslkog Yukawa texture at GUT scale induced by higher dimensional
operators. This model predicts lepton flavor violation. The predictive
features of the Georgi-Jarlskog texture are used to estimate branching ratios
for the radiative decays $e_{a} \rightarrow e_{b}+\gamma,$ and we find
rates that could provide an experimental test for this kind of model.
\end{abstract}
\end{titlepage}
\setcounter{page}{2}

\section {Introduction}

Supersymmetric Grand Unified Theory (SUSY GUT), supported by the unification of
the coupling constants, provides an interesting framework for extending the
Standard Model. In order to enhance  predictivity, several {\em ansatze}
for the Yukawa structure can be discussed in the context of  SUSY GUT models
inspired by superstrings. It is worthwhile to study whether some of the exact
predictions of the Standard Model (SM) are  modified in phenomenologically
interesting ways by
these proposals.

One of the most predictive {\em ansatze} for the Yukawa structure at GUT
is the  Georgi-Jarlskog \cite{GJ} texture,  which assumes symmetric quark and
lepton mass matrices of different forms for the up and down quark sectors.
Since in SO(10) all the fermions of a family are unified in an  irreducible
representation, the {\bf 16}, the symmetric texture is naturally
accommodated.  The \st representation can also
accommodate a  right handed neutrino, which leads to an interesting
phenomenology.

The generation of predictive textures with renormalizable
couplings for the Yukawa sector in SO(10) unified
models implies the introduction of large representations (like
$\overline{\bf 126}$) at about
the GUT scale.
This can destroy the asymptotic freedom of the theory; moreover, without
further input, the
hierarchies among the different couplings remain unexplained. In order to
deal with
both these issues,  it has been suggested that  the usual trilinear Yukawa
couplings be replaced by higher dimension operators \cite{FN} which
involve lower representations of SO(10). Recently, it has been shown that
such models can arise in the free  fermionic formulation of superstrings
\cite{Cha,Cle}. The authors of
these papers  classify specific SO(10)
representations which can emerge as  massless chiral multiplets below  the
Planck scale. Such constraints can be used to generate the Yukawa texture  at
GUT , and a particular example was provide in ref. \cite{Moh}. We will use the
model proposed in this reference in an illustrative manner to generate the
Georgi-Jarskog texture, and as an example for our calculations.

In the Minimal Supersymmetric Standard Model (MSSM), the terms which softly
break supersymmetry originate at the Planck scale. Thus, even with flavor
diagonal initial conditions at Planck scale, it has been observed
\cite{HKR,Mas} that the
evolution   between Planck and GUT scales may induce initial conditions in
these parameters at GUT scale which are not flavor-neutral, and  which
eventually generate important contributions to low energy flavor violation.
In this work we study lepton flavor-violating (LFV) proceses  that arise due to
the SO(10) structure  of the
superpotential above GUT.  We illustrate some
new important effects that originate with the requirement of the
Georgi-Jarlskog texture, and show that these effects could lead to LFV
lepton decay rates which are comparable
with the experimental limits.

In a recent  paper, Barbieri, Hall and Strumia \cite{Bar} have calculated
rates for LFV decays in the context of an SO(10) GUT. There is notable
difference between this present work and ref. \cite{Bar}: the large LFV effects
calculated in this work for
$\mueg$ processes do {\em not} depend on the  existence of
a large top Yukawa coupling - they are in fact completely decoupled from
the third generation. Instead, they are directly traceable to the
(above-stated) requirement of
generating  the Georgi-Jarlskog texture in an SO(10) GUT.

\section{Slepton Mass Matrix}

The minimal supersymmetric extensions of
the Standard Model (MSSM) coming from spontaneously broken
minimal N=1 Supergravity theories, can be described
by the superpotential \cite{HK}:
\begin{equation}
W = h_{ij}^{u} {Q}_{i}
{H}_{u} {U}_j + h_{ij}^{d}{Q}_{i}
{H}_{d} {D}_j + h_{ij}^{e}{L}_{i}
{H}_{d} {R}_j+
\mu {H}_{d} {H}_{u}\ \ ,
\end{equation}
where group indices have been omitted. In addition one introduces
all the allowed soft supersymmetry breaking (SSB) terms involving the scalars
and gauginos. These are given by
\cite{HK}
\begin{eqnarray}
-{\cal L}_{soft} &=&(\xi_{ij}^{u} Q_{i} H_{u} U_j + \xi_{ij}^{d} Q_{i}
H_{d} D_{j} + \xi_{ij}^{e}L_{i} H_{d} R_j +h.c)+
 (B \mu H_{d} H_{u}+h.c) \nonumber\\
& &
+m_{H_{u}}^2 |H_u|^2 + m_{H_{d}}^2 |H_d|^2+
m_{\tilde{L}}^2|\tilde{L}|^2+m_{\tilde{E}}^2|\tilde{E}|^2+
m_{\tilde{\nu}}^2|\tilde{\nu}|^2+
m_{\tilde{Q}}^2|\tilde{Q}|^2 \nonumber\\
& &
+m_{\tilde{D}}^2|\tilde{D}|^2+
m_{\tilde{U}}^2|\tilde{U}|^2
+\tfrac{1}{2} (M_{3} \tilde{\bar {g}} \tilde{g}+
M_{2} \tilde{\bar {W^a}} \tilde{W^a}+
M_{1} \tilde{\bar {B}} \tilde{B}+h.c)\ \ .
\label{eq:eta}
\end{eqnarray}
with $q=u,d,e$.

{}From the coupling of the MSSM to the minimal $N=1$ supergravity the
following set of assumptions is plausible at Planck scale:
\begin{eqnarray}
\label{eq:ic}
	M_{i}(\mp)&=&m_{1/2},\quad i=1,2,3\nonumber\\
	m_{\tilde{Q}}^2&=& m_{\tilde{u}}^2=...=m_{0}^2\ \ .
\end{eqnarray}
We make no statement about the $\xi_{ij}$ at Planck scale; boundary
conditions on $\xi_{ij}$ at GUT will follow from the physics to be
discussed.

It has been shown in refs. \cite{HKR,Mas} that in SUSY-GUT's, the
additional couplings to the extra GUT fields between GUT and Planck
scales can substantially modify the  universality of the SSB parameters
at GUT,  allowing
LFV  vertices. In our study we consider an
SO(10) SUSY-GUT model in which  the Yukawa textures are obtained at GUT from
effective higher dimensional operators. In this kind of model, it will be the
the different group structure of operators involved in the generation of the
Yukawa texture which produces a  flavor-dependent evolution of the SSB terms,
and which results in   LFV vertices.

As previously stated, we choose to illustrate our calculations using the model
of ref. \cite{Moh}. In  this model the Georgi-Jarlskog texture is derived from
the superpotential:
\begin{eqnarray}
\label{eq:moha}
W_{Yukawa} = M^{-1} \left( Y_{33}\psi_{3} \psi_{3} H_{1} S_{1}+
h_{33}\psi_{3} \psi_{3} H_{2} S_{2}+h_{23}\psi_{2} \psi_{3} H_{2} S_{3}
\right) \nonumber\\
+M^{-2} Y_{22}\psi_{2} \psi_{3} A A\pri H_1+
M^{-3} \left( Y_{12}\psi_{1} \psi_{2} H_{1} S_{2}^3+
h_{12}\psi_{1} \psi_{2} H_{2} S_{1}^3\right)
\end{eqnarray}

This superpotential is SO(10) invariant,  and respects a set of discrete
symmetries (given in \cite{Moh}).
\ $M$ is an intermediate mass between Planck and GUT that could arise by
integrating out fermions like $\overline{\st} +\st$.\ We will take
$M=\mp=2.4\times 10^{18}\gev.$ The fields $S_1,S_2,S_3$
are singlets,\ $H_1,H_2$ are {\bf 10}'s,\ and $ A,A\pri$ {\bf 45}'s of SO(10).
It is important to note that there are {\em no renormalizable interactions}
in the Yukawa sector.

In the 2-2 entry,\ the operator $H_1 A A'$ generates the
$\overline{\bf 126}$  required by the Georgi-Jarlskog texture.\ The vevs of
$A$ and $A\pri$ are along the $(B-L)$ and the $I_{3R}$ directions,
respectively, so only  the $\overline{\bf 126}$ contributes to the mass
matrices~\cite{Moh}.

The couplings in the superpotential are chosen to be of order 1. The
vevs $\vev{S_1},$ $\vev{S_3},$ $\vev{A},$ $\vev{A\pri}$ are supposed
to be of the order of
the GUT scale,\ $\vev{S_2}
\approx M$ and we assume for simplicity that all of them are real.

As follows from the standard supergravity analysis \cite{Sug},
soft-breaking
terms (the equivalent of the usual trilinear term) of the form
\be
-{\cal L}^{10}_{soft}= M^{-1}  \overline Y_{33}\psi_{3} \psi_{3} H_{1} S_{1}+
\ldots
\label{eq:soft}
\ee
are generated from (\ref{eq:moha}). Matching to Eq.~(\ref{eq:eta}) at GUT, we
have at tree level relations such as
\be
(\xi^{d,e}_{33})\sg=M^{-1}(\overline Y_{33})\sg\vev{S_1}\ \ .
\label{eq:baryxi}
\ee

Then we assume that at GUT all the symmetries break at once to
$SU(3)_c\times SU(2)_L\times U(1)_Y,$ generating the  effective Yukawa texture
\begin{equation}
\label{eq:GJu}
h^u=\left(\begin{array}{ccc} 0&C&0 \\
C&0&B\\
0&B&A
\end{array}\right)
\end{equation}

\begin{equation}
\label{eq:GJd}
h^d=\left(\begin{array}{ccc} 0 & F & 0 \\
F & E & 0\\
0 & 0 & D
\end{array}\right)
\end{equation}

\begin{equation}
\label{eq:GJe}
h^e=\left(\begin{array}{ccc} 0 & F & 0 \\
F & -3 E & 0\\
0 & 0 & D
\end{array}\right)
\end{equation}
If the vevs satisfy $\vev{S_1}\sim\vev{S_3}\sim\vev{A}\sim\vev{A\pri}\sim \eps
M, \vev{S_2}\sim M,$ then a hierarchy among the Yukawa entries follows:
\[
A=h_{33};\ B=h_{33} \epsilon;\ C=h_{12} \epsilon^3;\ D= \epsilon Y_{33};\
E=\epsilon^2 Y_{33};\ F=\epsilon^3 Y_{33}\ .\]
Even though $\vev{S_2}/M\sim 1,$  we treat the 33 piece of the lagrangian as
nonrenormalizable.

For superpotentials like the one in Eq.(\ref{eq:moha})
the universality of the initial conditions (\ref{eq:ic}) is broken at GUT;
the soft masses evolve differently according to their representation $\rc$
under
SO(10), reaching GUT values
\begin{equation}
\label{eq:m16}
m_{\rc}^2=m_{0}^2+2 C_{\rc}\ \mg^2\ \atg\left[\tp(2 -b_{10}
\atg \tp)/(1-b_{10} \atg \tp)^2\right]\ \ .
\end{equation}
$C_{\rc}$ is the Casimir for the
representation $\rc\  (\rc={\bf 16, 10}),$ and
$b_{10}=4$. We define $\atg=\alpha_G/4\pi$, and the
variable
$t=2\log(Q/\mgut),$ with $\tp=2\log(\mp/\mgut).$
$\mg$ is the gaugino mass at GUT, related
to $m_{1/2}$ by:
\[
\mg=(1-b_{10} \atg \tp)\ m_{1/2}\ \ .
\]

Radiative corrections coming from (\ref{eq:moha}) must also be included in
obtaining a slepton mass matrix at GUT; as we shall see, these will
substantially alter the flavor-diagonal expression (\ref{eq:m16}) in the case
of $\rc={\bf 16}$.

The charged slepton mass matrix can be written at Fermi scale as:
\begin{eqnarray}
-{\cal L}_m^{sl}&=&
\tilde{e^\dagger}_{L}\left(m_{L}^2+\delta m_{16}^2+m_{e}
m_{e}^+\right)\tilde{e}_{L}+\tilde{e^\dagger}_{R}\left(m_{R}^2+\delta
m_{16}^2+m_{e}
m_{e}^+\right)\tilde{e}_{R}\nonumber\\
& &+\tilde{e^\dagger}_{L}\left(
\frac{v_{d}}{\sqrt{2}}(\xi^e
+ \delta\xi^e)+\mu m_{e} \tan\beta\right)\tilde{e}_{R} + h.c.
\end{eqnarray}
where all entries are matrices in flavor space. In a standard notation,
$\tan\beta=v_u/v_d.$
$m_{e}$ is the charged fermion  mass matrix, and  $\delta m_{16}^2$ and $\delta
\xi^e$ are additional contributions generated via radiative corrections
between $\mp$ and
$\mgut$ which are sources of flavor violation;
they will be explicitly given in
what follows.
$m_{L}^2$ and $m_{R}^2$ are flavor independent and are
given to one loop by the expressions:
\begin{eqnarray}
m_{L}^2&=&m_{16}^2 +\left(\tfrac{3}{10}K_{1}(t)+\tfrac{3}{2}K_{2}(t)
\right) \atg \ \mg^2\nonumber \\
&&\hspace*{5cm}-m_{Z}^2 \left(\tfrac{1}{2}-\sin^2\theta_W
\right) \cos 2\beta\label{eq:ml}\\
m_{R}^2&=&m_{16}^2 +\tfrac{6}{5}K_{1}(t)
\cdot \atg\ \mg^2 -m_{Z}^2 \sin^2\theta_W \cos 2\beta
\label{eq:mr}
\end{eqnarray}

We have defined $K_{i}(t)=t (2+b_{i} \atg t)/(1+
b_{i} \atg t)^2$ ,\ with $b_i=(33/5,1,-3)$ and
$\theta_W=$Weinberg angle. $m_{16}^2$ is given in Eq.~(\ref{eq:m16}).
The mass
matrix for the (left-handed) sneutrinos is given by:
\begin{equation}
(\tilde{\nu}^+_L) \left(
m_{16}^2 +\left(\tfrac{3}{10}K_{1}(t)+\tfrac{3}{2}K_{2}(t)
\right) \atg \mg^2 + \tfrac{1}{2}m_{Z}^2  \cos 2\beta
+\delta m_{16}^2 \right)(\tilde{\nu}_L)
\end{equation}


\section {Non-universal Soft Scalar Masses At GUT}

The entries for the above-GUT correction  $\delta m_{16}^2$ come from
one loop diagrams like the one in Fig 1, and can be generated by
evolving the composite operators from Planck to GUT. We illustrate our
procedure by considering the 2-3 entry. The interactions required for
calculating this one-loop contributions can be obtained from the
relevant $F$-terms
\begin{eqnarray}
F_{H_2} F_{H_2}^{\ast}
&=&\left(\frac{h_{23}}{M} \right)\left(\frac{h_{33}}{M}\right) \psi_3
\psi_{3}^{\ast}
\psi_3
\psi_{2}^{\ast} S_2 S_{3}^{\ast}+ h.c\nonumber\\
F_{\psi_3} F_{\psi_3}^{\ast}&=&
2 \left(\frac{h_{23}}{M} \right)\left(\frac{h_{33}}{M}\right) \psi_3 H_2
H_2^\ast \psi_2^{\ast} S_2  S_3^{\ast}+ h.c\nonumber
\end{eqnarray}
 Its effective
contribution at GUT can be estimated by integrating the loop between the two
scales, so that
\begin{equation}
(\delta m_{16}^2)_{23}=-2\cdot 5\ \frac{m_{16}^2}{8
\pi^2}\ \log\frac{\mp}{\mgut}\ (2 h_{23} h_{33})\frac{\vev{S_2}}{M}
\frac{\vev{S_3}}{M}
\end{equation}
The factor of 5 counts the fields of the multiplet runing in the loop
and is the same for the \st\ and for the {\bf 10}. \ There is  an
additional symmetry  factor of   2 when $\psi_3$ runs in the loop.
In this fashion we find for the complete matrix  $\delta
m_{16}^2$
\begin{equation}
\label{eq:del1}
\delta m_{16}^2=-10\ \left(\begin{array}{ccc}
y_{12}^2 & 0 & BC \\
0 & y_{12}^2 & 2 BA\\
BC & 2 A B  & 4 A^2
\end{array} \right) \frac{m_{16}^2}{8 \pi^2}
\ \log\frac{\mp}{\mgut}
\end{equation}
The 0 value for the 1-2 entry is due to the orthogonality of the vevs of
$A, A\pri.$ \ We have displayed only  the lowest power of $\epsilon$ in every
matrix element, which  is equivalent to neglecting the down Yukawas.

We pause for two remarks concerning Eq.~(\ref{eq:del1}):
\bi
\item In models with renormalizable couplings, where
the hierarchy is enforced by
having different Higgs fields to generate the Yukawa entries, there
will be no off-diagonal elements induced in $\delta m^2$ at one-loop level.
This
contrasts with our result, as evident in Eq.~(\ref{eq:del1}).
\item
The third generation receives a large contribution $(\delta m_{33}^2)$ from
the top Yukawa A. (In our model, $A\approx 3.$) This leads to a significant
(indeed, nonperturbative) splitting of $m_{\tilde{\tau}}^2$ from
$m_{\tilde{\mu}}^2\simeq m_{\tilde{e}}^2$. (This possibility was noted in
ref. \cite{Bar}). For $\vev{S_2}/M\sim 1,$ there are other possible
contributions to
$(\delta m_{33}^2)$ \cite{Non}. In
the present model, flavor violations in the
$\mu-e$ sector will not involve the $\tilde{\tau}$ or its mass. Flavor
violations in
$\tau-\mu$ or
$\tau-e$ processes will involve loops with $\tilde{\tau},$ and this mass
splitting will be simply parameterized when these processes are discussed.
\ei

Both the coupling constants in the superpotential (\ref{eq:moha})  and the
related soft-breaking parameters (\ref{eq:soft}) receive radiative
corrections in evolving from Planck to GUT. These are
obtained by integrating the
loops in Fig 2 between the two scales (we display only the 22 entry). Displayed
as differential equations, the results
are
\beq
\frac{d Y_{ab}}{dt}& = & 2 G_{ab}\at_{10} Y_{ab}\\
\frac{d \overline Y_{ab}}{dt}& = & -2 G_{ab}\at_{10}(2 M_{10} Y_{ab}
-\overline Y_{ab})\ \ ,
\label{eq:rg}
\eeq
where $\at_{10}$ is the (running) SO(10) coupling constant (divided by $4\pi)$.
and $M_{10}$ is the (running) SO(10) gaugino mass. Their behavior with scale
$t$ is governed by the usual SO(10) R-G equations, appropriate to the
representation content of the model of ref. \cite{Moh}.

The factors $G_{ab}$ depends on the group structure of the Yukawa operator,
and one can show that they are
\begin{equation}
 G_{ab}=-\frac{1}{2}\sum_{r}C_r
\end{equation}
where $C_r$ is the Casimir operator for the representation of every field
present in the operator.
An  integration of Eq.(\ref{eq:rg}) using Eq.~(\ref{eq:ic}) and the
matching conditions (\ref{eq:baryxi}) gives for the
effective trilinear parameter
at GUT
\begin{equation}
(\xi_{ab}^e)_{G}=m_0 A_{0}( h_{ab}^e)_{G}+8\atg\ G_{ab}\ m_{1/2}
 (h_{ab}^e)_{G} \log\left(\mp/\mgut\right)
\end{equation}
where the subscript $G$  denotes evaluation at GUT. We have used as a boundary
condition $(\overline Y_{ab})_{\mp}=m_0A_0(Y_{ab})_{\mp}.$

In our example the factors $G_{ab}$ are $-63/8$ for all $\{ab\}$ except for
$\{ab\}=\{22\},$ in which case it is $-127/8.$ Matching our results at GUT to
the MSSM  lagrangian, the expressions for the trilinear soft terms for the
sleptons  becomes:

\[
(\xi_{ab}^e)_{G}=m_0 A_{0}\left(\begin{array}{ccc} 0 & F & 0 \\
F & -3 E & 0\\
0 & 0 & D
\end{array}\right)\hspace{5cm} \]
\be
+8\atg\ m_{1/2}\
  \log\left(\frac{\mp}{\mgut}\right)
\left( \begin{array}{ccc}
0&(-63/8)F&0\\
(-63/8)F&(-127/8)(-3E)&0\\
0&0&D \end{array}\right)
\label{eq:etag}
\ee


In what follows, it is convenient to work in a superfield basis in which the
lepton Yukawa matrix is diagonal at GUT. In this basis, only $\delta m_{16}^2$
and $\delta\xi^e$ mix the generations. Also in this basis, no new off-diagonal
entries are generated during the evolution from GUT to Fermi scale, so that
only
the terms
$\delta m_{16}^2$ and $\delta\xi^e,$ containing GUT-scale physics, contribute
to lepton flavor mixing. We also note here the different nature of the mixing
between different generations: $\delta m_{16}^2$ is responsible for the mixing
of the third  generation with the first two, and this mixing is permitted
through the use of non-renormalizable operators above GUT.
In models where the G-J texture arises from renormalizable operators
(including Higgses in {\bf 10}'s and $\overline{\bf 126}$'s), the
Higgs stucture chosen to enforce the desired Yukawa texture at GUT prevents the
mixing of the third generation with the others in the slepton matrix  at GUT.
Enhancement in the mixing between the first two generations (in $\delta \xi^e$)
comes from sizeable SO(10) group theoretic factors.

 We denote
with a  Greek index the lepton mass eigenstates, and rotate  the superfields:
\begin{eqnarray*}
\widehat{e}_{R_\alpha}&=&U_{\alpha j}^R\ \widehat{e}_{R_j}\\
\widehat{e}_{L_\alpha}&=&U_{\alpha j}^L\ \widehat{e}_{L_j}
\end{eqnarray*}
such that
\begin{equation}
h_{Diag}^e=U^L h^e U^{R\dagger}
\end{equation}
Since $ h^e$ is symmetric,\ up to a phase we can identify $U^L$ and $U^R$
with a matix $U_e$ such that $U_e h^{e\dagger}h^e
U_e^{\dagger}=(h_{diag}^e)^2$.\ In  the Georgi-Jarskog texture this matrix is
of the form:
\begin{equation}
\label{eq:U}
U_{e}=\left( \begin{array}{ccc}\cos\theta & -\sin\theta & 0\\
		               \sin\theta & \cos\theta & 0 \\
				0 & 0 & 1 \end{array}
\right)
\end{equation}
where
\begin{equation}
\tan\theta =-\frac{2 F}{3 E}
\end{equation}

Then the non-diagonal terms in the slepton matrix are contained in:
\begin{eqnarray}
\Delta^{LL}&=&\Delta^{RR}=U_{e}\ \delta m_{16}^2\
U_{e}^{\dagger}\label{eq:del13}\\
\Delta^{LR}&=& U_{e}\ \frac{v_{d}}{\sqrt{2}}\ (\delta\xi^{e})\sg\ U_e^{\dagger}
\label{eq:dlr}
\end{eqnarray}
This makes explicit the property that all flavor mixing
in the slepton sector is a
reflection of physics above GUT.
 Using expressions (\ref{eq:del1}), (\ref{eq:U})
we find for the mixing of the third generation:
\begin{eqnarray}
\Delta_{\tau e}&\equiv&\Delta_{13}^{LL}=\Delta_{13}^{RR}=-(BC\cos\theta
-BA\sin\theta)\ 10\
\frac{m_{16}^2}{8 \pi^2}\ \log\left(\frac{\mp}{\mgut}\right)
\label{eq:in13}\\
\Delta_{\tau \mu}&\equiv&\Delta_{23}^{LL}=\Delta_{23}^{RR}=-(BC\sin\theta
+2BA\cos\theta)10
\frac{m_{16}^2}{8 \pi^2}\log\left(\frac{\mp}{\mgut}\right)
\label{eq:in23}
\end{eqnarray}
{}From the trilinear terms we find (using (\ref{eq:etag})\ and
\ (\ref{eq:dlr}))
for the  mixing of the first two generations:
\begin{equation}
\label{eq:delr}
\Delta_{\mu e}\equiv\Delta_{12}^{LR}=F\frac{v_{d}}{\sqrt{2}} \cos2\theta
(-63/8+127/8)\  8 \atg\ m_{1/2}\ \log\left(\frac{\mp}{\mgut}\right)
\end{equation}
We can see that in our model:
\[
F\cdot v_d\approx \sqrt{m_e m_{\mu}},\ \ E\approx m_{\mu}\ \ .
\]
Then (\ref{eq:delr}) is of the order of
\[
\sqrt{m_e m_{\mu}}\cdot m_{1/2} \cdot Clebsch
\]
This result can be compared with the effective insertion obtained by Barbieri
{\em al} \cite{Bar}: in their model (which does not contain the G-J texture)
\[
\Delta_{12}^{LR} \approx m_{\tau} V_{\tau \mu} V_{\tau e} m_{1/2}
\]
Parametric agreement with  our result is obtained for
\[
V_{\tau \mu} V_{\tau e}\approx\sqrt{\frac{m_{\mu}}{m_{\tau}}}
\sqrt{\frac{m_e}{m_{\tau}}}
\]
which is the approximate situation for the model in \cite{Bar}.
This explains why using a different approach our results are numerically
comparable with theirs.

\section {Radiative Decay $e_a\rightarrow e_b\ \gamma$}

The amplitude for the decay can be written as a magnetic transition

\begin{equation}
T(e_a\rightarrow e_b\gamma)=\epsilon^{\lambda}
{\bar e}_b(k-q)\ [ iq^\nu \sigma_{\lambda \nu}(A+B\gamma_{5})]\ e_a(k)
\label{eq:mag}
\end{equation}
In the limit of vanishing mass for the outgoing leptons the left handed and
right handed decay amplitudes do not interfere therefore Eq.~(\ref{eq:mag}) can
be written as:
\begin{equation}
\label{eq:amp}
T(e_a\rightarrow e_b\gamma)=\epsilon^{\lambda}
{\bar e}_b(k-q) \ \lbrace 2k\epsilon\lbrack A_{R}
\left(\frac{1+\gamma_5}{2}\right)+
A_{L}\left(\frac{1-\gamma_5}{2}\right)\rbrack\ \rbrace\ e_a(k)
\end{equation}
Thus the decay rate is given by:
\begin{equation}
\Gamma(e_a\rightarrow e_b\gamma)=\frac{{m_{e_a}}^3}{16\pi}
(|A_R|^2+|A_L|^2)
\label{eq:decay1}
\end{equation}
Since the mass insertions depend on the generations, the  diagrams $(a,b,c)$
of
Fig 3 contribute to \taueg\ and
\taumg,
while the diagram of Fig 4 contributes to \mueg. For the
$\tau$ decays we obtain
\begin{equation}
A_{L}=A_{L}^{3b};\qquad \ A_{R}=A_{R}^{3a}+A_{R}^{3c}\ \ ,
\end{equation}
where
\begin{eqnarray} A_{R}^{3a}&=&i \frac{g^2 e}{32 \pi^2 \cwt}
\frac{\Delta_{\tau \mu} m_{\tau}}{(m_{\tilde{\tau}_L}^2-m_{\tilde{\mu}_L}^2)}
\frac{F_{3a}}{m_{{\tilde\chi_j}^0}^2}\\
A_{L}^{3b}&=&i \frac{g^2 e}{32 \pi^2 \cwt}
\frac{\Delta_{\tau \mu} m_{\tau}}{(m_{\tilde{\tau}_R}^2-m_{\tilde{\mu}_R}^2)}
\frac{F_{3b}}{m_{{\tilde\chi_j}^0}^2}\\
A_{R}^{3c}&=&-i \frac{g^2 e}{16 \pi^2 }
\frac{\Delta_{\tau \mu} m_{\tau}}{(m_{\tilde{\nu}_{\tau}}^2-
m_{\tilde{\nu}_{\mu}}^2)}
\frac{F_{3c}}{m_{{\tilde\chi_j}^{+}}^2}\ \ ,
\end{eqnarray}
while for \mueg
\begin{equation}
A_{R}^{4a}=A_{L}^{4b}=+i \frac{g^2 e}{32 \pi^2 \cwt}\frac{\Delta_{\mu e}^{LR}}
{m_{\tilde{\mu}_L}^2-m_{\tilde{e}_R}^2} \frac{F_{4}}{m_{\tilde\chi_j^0}}\ \ .
\end{equation}

We have defined:
\begin{eqnarray} F_{3a}&=&\left|\cw Z_{2j}+\sw Z_{1j}\right|^2 \left[
g\left(m_{\tilde{\tau}_L}/
m_{\tilde\chi_j^0}\right)-g
\left(m_{\tilde{\mu}_L}/m_{\tilde\chi_j^0}\right)
\right]\\ F_{3b}&=&|2 \sw Z_{1j}|^2  \left[ g\left(m_{\tilde{\tau}_R}/
m_{\tilde\chi_j^0}\right)-g
\left(m_{\tilde{\mu}_R}/m_{\tilde\chi_j^0}\right)
\right]\\ F_{3c}&=&|V_{1j}|^2 \left[ f\left(m_{\tilde{\nu}_{\tau}}
/m_{\tilde\chi_j^+}\right)-f
\left(m_{\tilde{\nu}_{\mu}}/m_{\tilde\chi_j^+}\right)
\right]\\ F_{4}&=&(\cw Z_{2j}^\ast+\sw Z_{1j}^\ast)(2 \sw Z_{1j}^\ast)
\left[ h\left(m_{\tilde{\mu}_L}/
m_{\tilde\chi_j^0}\right)-h
\left(m_{\tilde{e}_R}/m_{\tilde\chi_j^0}\right)
\right]
\end{eqnarray}
where $Z_{ij}$ and $V_{ij}$ are  the neutralino and chargino
mixing matrix  respectively defined as in Haber and Kane \cite{HK},
 and

\begin{eqnarray}
\label{eq:int}
f(a)&=&\frac{2 +3 a^2 -6 a^4+ a^6+6 a^2\log(a^2)}{12 (a^2-1)^4}\\
g(a)&=&\frac{5- 9 a^4 +4 a^6 +6 a^2 [2 \log(a^2) -a^2 \log(a^2)]}{(a^2-1)^4}\\
h(a)&=&\frac{1 -a^4 +2 a^2 \log(a^2)}{2 (a^2-1)^3}
\end{eqnarray}

Eq.~(\ref{eq:decay1}) gives for the  ratio
\begin{equation}
\frac{\Gamma(e_a\rightarrow e_b \gamma)} {\Gamma(e_a\rightarrow
e_b \nu_a\bar{\nu}_b)}=
\frac{12 \pi^2}{m_{e_a}^2 G_{F}^2}(|A_R|^2+|A_L|^2)
\end{equation}
Since the percentage of the total decay for the reactions in the
denominator  are $99\%$ for \mueg, $\ 18.01\%$ for
\taueg, $\ 17.65\%$ for
\taumg\ \cite{Dat}, we write for the
relevant branching ratios:
\begin{eqnarray}
\mbox{BR}(\mueg)&=&\frac{3 \alpha}{\pi} \left(
\frac{\Delta_{\mu e}}{m_{\tilde\mu_L}^2-m_{\tilde e_R}^2}\right)^2
\frac{M_Z^4}{m_{\mu}^2 m_{\tilde\chi_j^0}^2}\
|F_4|^2\hspace*{3cm}\label{eq:br12}\\
\mbox{BR}(\taumg)&=&\frac{3 \alpha}{2 (.18) \pi}\
M_{Z}^4\ (\Delta_{\tau \mu})^2 \quad\cdot\nonumber
\end{eqnarray}
\begin{equation}
\left( \left|\frac{F_{3a}}{(m_{\tilde{\tau}_L}^2-
m_{\tilde{\mu}_L}^2)m_{\tilde\chi_j^0}^2}  -2
\cwt\frac{F_{3c}}{(m_{\tilde{\nu}_{\tau}}^2-m_{\tilde{\nu}_{\mu}}^2)
m_{\tilde\chi_j^+}^2}\right|^2 + \left|\frac{F_{3b}}{(m_{\tilde{\tau}_R}^2-
m_{\tilde{\mu}_R}^2)m_{\tilde\chi_j^0}^2}\right|^2\right)
\label{eq:br23}
\end{equation}
Since in our model the slepton (mass)$^2$  of the two first generations differ
only in  the square of the lepton masses, we can use Eqs.~(\ref{eq:in13})
and (\ref{eq:in23})  to obtain
\begin{equation}
\label{eq:br13} \mbox{BR}(\taueg)=
\left(\frac{\Delta_{\tau e}}{\Delta_{\tau \mu}}\right)^2 \cdot
\mbox{BR}(\taumg)
\end{equation}

\section {\bf Inputs}

In order to estimate the branching ratios of expressions
(\ref{eq:br12}),(\ref{eq:br23}),(\ref{eq:br13})
we require
values for the GUT coupling $\alpha_{\rm G},$ the common scalar mass at
Planck scale
$m_0^2,$ the gaugino mass at Planck scale  $m_{1/2},$ the (effective) Higgs
bilinear coupling at GUT $\mu,$ the ratio of vevs $\tan\beta,$ the
flavor-symmetric soft-breaking parameter   $A_0,$ and the
Yukawa matrices at GUT.

The value of $\mu$ up to its sign can be expressed in terms of the other
parameters by means of the symmetry breaking  relation:
\begin{equation}
\label{eq:mu}
\mu^2=-\frac{m_{Z}^2}{2}-\frac{m_{H_d}^2-m_{H_u}^2 \tan^2{\beta}}{1-
\tan^2{\beta}}
\end{equation}
In this expression we neglect the one loop corrections to the the
effective  potential,\ although these could be significant for some values of
the  parameters $m_{1/2},m_{0},$ and $\tan\beta $ \cite{AN}. Its inclusion
unnecessarily complicates our calculations,\ since  in our case the branching
ratios are {\em not} proportional to $\mu$ or  to $A$ (unlike the situation in
ref. \cite{Bar}). In our case  $\mu$ is primarily used in determining the mass
matrix for the neutralinos.

To obtain $\alpha\sg$ and $\mgut$ we integrate the R-G equations
assuming  for integration purposes that all the supersymmetric masses, as well
as the heavier Higgs, are   degenerate at $m_t.$
We consider the light Higgs to have a mass of the order of $M_Z$.\ On
integrating the MSSM to two loops in the coupling constants \cite{MV} from
GUT to $m_t$ and the SM from there  to
$M_Z,$ we find
\[ \alpha\sg = .041;\ \ \mgut=2\times 10^{16}
\]
obtained for the three coupling constants at $M_Z$
\[
\alpha_{s}(M_Z)=.120;\ \
\alpha_{1}(M_Z)=.0169;\ \ \alpha_2(M_Z)=.0332
\]
The value of $A_{0}$ is used in the integration of the R-G equations to obtain
$\mu.$ Varying it from $-3$ to $3$ produces a change of less than the
$5\%$ in  $\mu$; in the range of  approximation we are using we can
fix $A_0=1$ for the rest  of the calculation. As we said before
the branching ratios are not  proportional to this parameter.

There are in the literature several studies of the predictivity of the
Georgi-Jarlskog texture \cite{Dim2,BB}. In these, the
Yukawa matrices are scaled from GUT to Fermi scale,
giving  reasonably good agreement with the
experimental data. We concentrate on low values of $\tan \beta$ $(\ <10 \ )$ so
we can assume  that
$h_{t} >> h_{\tau},h_{b}$ and use the one loop semianalytical analysis of
ref. \cite{Dim2} to determine our imputs. The predictivity of
this model is still
impressive although the value obtained for $V_{cb}$ is at the upper limit of
the
most recent  experimental values \cite{Neu}.  We follow here the one  loop
analysis as in \cite{Dim2}.

We use as inputs
\[ m_\tau=1.784\gev;\ m_e=.511\mev;\ \ m_{\mu}=105.658\mev;\]
\[m_u/m_d=.55 ;\ m_b=4.23\gev; \ m_c=1.26\gev\ \]
Working to three-loop
corrections in QCD
\cite{Go} and one loop  in QED we find
$h_t(m_t)=1.11$ for $A=3.08$ (where $A$ is
defined in Eq.~(\ref{eq:GJu})). For $1.6 \leq \tan{\beta} \leq 10,$
we obtain $.0446 \leq V_{cb} \leq .048.$
These lie at  the upper limit of the
experimental data.
The
physical mass for the top lies in the range $174\gev \leq m_t^{phy} \leq
204\gev,$ consistent with experimental bounds.

For the G-J texture (at GUT) we find:
\[ A=3.08;\ \ \ B=.094 (\sin\beta)^{-1/2};\ \ \ C= 1.67\ 10^{-4}
(\sin\beta)^{-1}; \]
\[ D=6.84\ 10^{-3} (\cos\beta)^{-1};\ \ \ E=1.35\ 10^{-4}
(\cos\beta)^{-1}; \]
\[ F=2.82\ 10^{-5} (\cos\beta)^{-1}\ \ . \]
We observe that the large  value of obtained for $A$  makes the
$\delta m^2$ radiative  correction (\ref {eq:del1})  for the third generation
highly non-perturbative (although the corrections to the Yukawa coupling
remain perturbative for this value of $A$ \cite{BB}). We deal with this problem
in the following manner: we consider as degenerate the
first two generations of sleptons,  and describe the reduction of the third
generation  slepton masses with a parameter $x$ as follows:
\begin{eqnarray}
m_{\tilde{\tau}_{L(R)}}^2&=&m_{\tilde{\mu}_{L(R)}}^2
-(1-x)m_{16}^2+m_{\tau}^2 \nonumber\\
m_{\tilde{\nu}_{\tau}}^2&=&m_{\tilde{\nu}_{\mu}}^2-(1-x)m_{16}^2
\label{eq:x}
\end{eqnarray}
This is equivalent to taking $m_{\tilde{\tau}_{L(R)},\
\tilde{\nu}_{\tau}}^2|_{\rm GUT}=x\ m_{16}^2,\ \ m_{\tilde{\mu}_{L(R)},\
\tilde{\nu}_{\mu} }^2|_{\rm GUT}=m_{16}^2.$


\section{\bf Results and Conclusions}

In Figures 5 and 6 we show the branching ratios obtained for
representative values of the input parameters.\ We use as one parameter
$m_{\tilde{\mu}_L}$ at the scale of $m_t$ (related to the soft mass for the
{\bf
16} at GUT by eq. (\ref{eq:ml})). We consider for
$\mg$ (the gaugino mass at GUT) a range from $50\gev$ (the approximate lower
limit from direct chargino search) to a maximum value that we consider to be
the
one that drives to zero
the mass of the scalars  in the  {\bf 10} at GUT,\ using
Eqs.~(\ref{eq:m16}) and (\ref{eq:ml}).

To illustrate our calculations, we have shown results for  $\tan\beta=3$.  The
change  with $\tan\beta$ is not very significant: the
branching ratios decrease slowly when we increase $\tan\beta$ to
its limit value of 10.

A change of  sign in $\mu$ is shown in Fig 5  to have little  effect on the
rates for \mueg. The same is true for \taumg.
The difference of the masses of the third generation of scalars
at GUT,\  represented by the parameter $x,$ only indirectly (and negligibly)
affects the rate for \mueg, by modifying  the value of
$\mu$. The effects on  \taumg\  are more direct, as
described in the text, and they are displayed in Fig 6 for
$m_{\tilde{\mu}_L}=300\gev$. (In the other curves we keep
$x=.5$.)

The results for $\mbox{BR}(\taueg)$ can be obtained by scaling the
ones of \taumg\  by a factor of $4.8\ 10^{-3}$   for
$\tan\beta=3$ ,\ (see Eq.~(\ref{eq:br13})).\ The rates we find are  obviously
much lower than the experimental limits.

In Fig 7 we show the areas of the plane $(m_{\tilde{\mu}_L},\mg)$ restricted by
our calculations. Already, certain regions of the presently allowed MSSM
parameter space are ruled out by the the present experimental limits.  We
observe that if the current limits were decreased by a factor of 10, a  large
portion of the  slepton mass range of phenomenological interest would be
excluded. Conversely, either or both of \mueg\ and
\taumg\  could be observed for superpartner masses in
the few hundred GeV range at the price of a factor of 10 improvement in
experiment.

To conclude: we have analyzed some
phenomenological consequences of embedding the MSSM in SO(10) GUT models
in which  Yukawa structures are generated by effective composite operators
at and above GUT.  The use of the Georgi-Jarlskog texture
permits the  use the low energy data to minimize the number of free
parameters, but also imposes  important constraints on the GUT physics.
In our analysis we have shown how a model which incorporates the G-J
texture predicts flavor-violating  phenomena that can be
tested in the current or in the next generation of experiments on lepton
decays.
In the case of $\mueg,$ the large enhancement of the rate originates with the
group structure associated with the presence of  an effective $\overline{\bf
126}$ in the higgs sector; we then expect our result to be valid in any SO(10)
model which incorporates the Georgi-Jarlskog texture.

\section*{Figure Captions}
Figure 1: One loop contribution to the 3-2 entry of $\delta m_{16}^2.$
\bigskip

\noindent Figure 2.1: One loop contributions to the effective trilinear\
$(\xi)$\ term at GUT: 2.1a, 2.1b, 2.1c  are  gauge, D-term, and gaugino
contributions, respectively.\\
\noindent Figure 2.2: One loop contributions to the effective Yukawa.
\bigskip

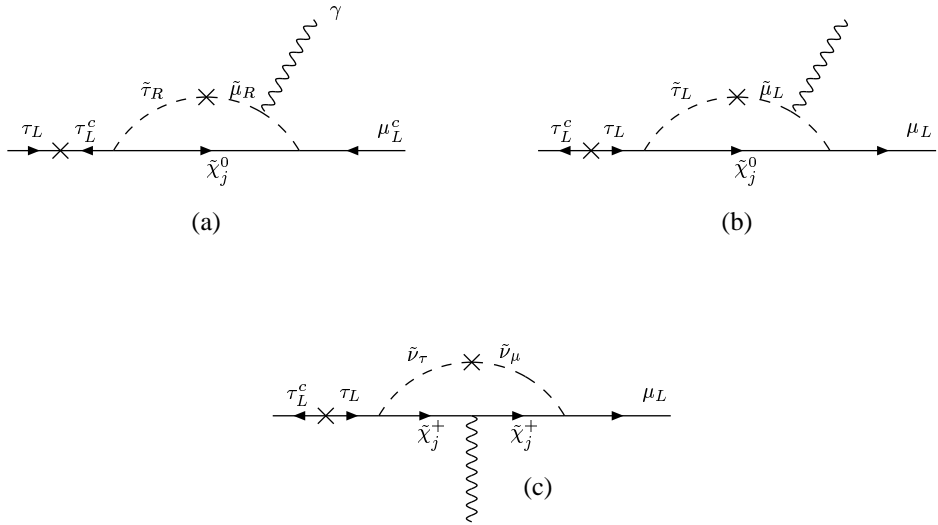
\noindent Figure 3: Diagrams contributing to \taumg. Diagrams similar to
3a, 3b with the photon line attached to $\tilde\tau$ are not displayed.
\bigskip

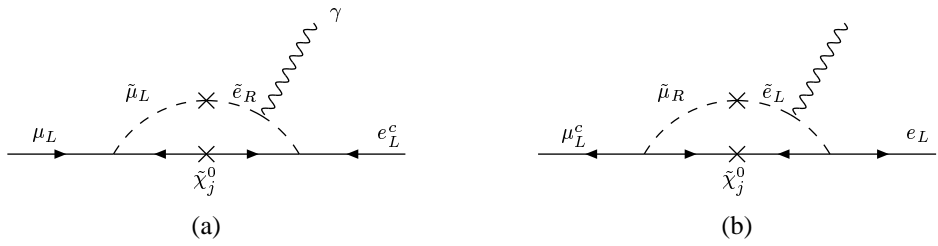
\noindent Figure 4: Diagrams contributing to \mueg. Diagrams similar to
4a, 4b with the photon line attached to $\tilde\mu$ are not displayed.
\bigskip

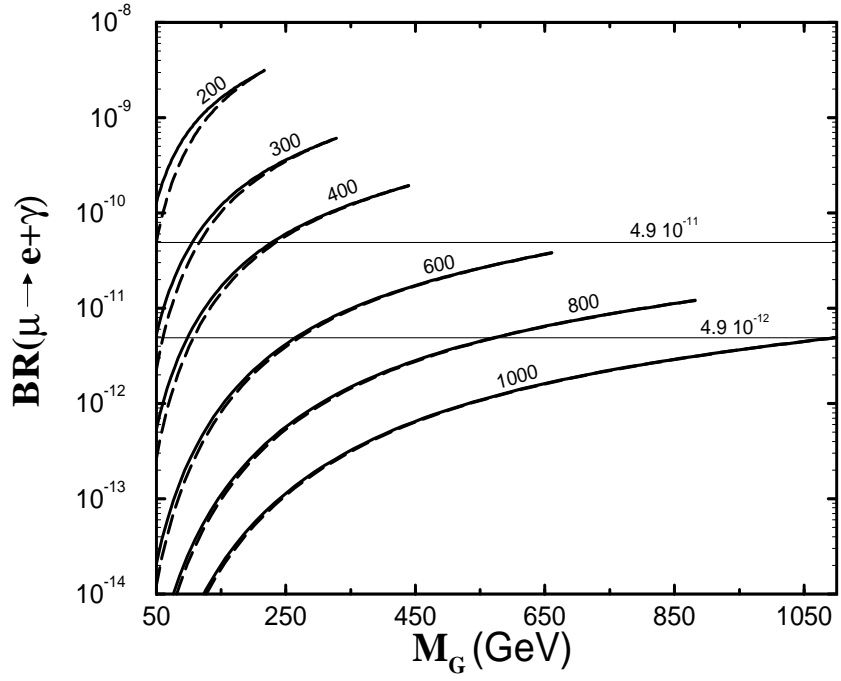
\noindent Figure 5: BR$(\mueg)$ for a range of values of
$m_{\tilde\mu_L}$ (labeling the curves) and gaugino mass at GUT
$(\mg).$ All curves are for $\tan\beta=3$ and $x=1$ (see Eq.~(\ref{eq:x}).
Solid
line:
$\mu>0;$ dashed line $\mu<0.$
\bigskip

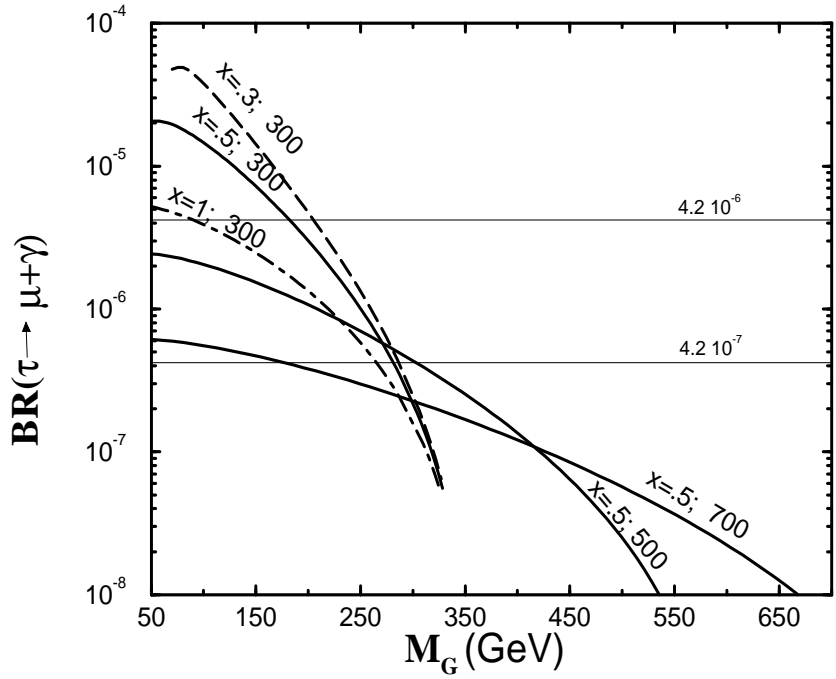
\noindent Figure 6: BR$(\taumg)$ for a range of values of
$m_{\tilde\mu_L}$ (labeling the curves) and gaugino mass at GUT $(\mg).$ All
curves are for $\tan\beta=3,\ \mu>0.$ The value of $x$ (the
$m_{\tilde\tau}^2$ suppression factor at GUT) has been chosen to be 0.5 for the
solid lines. The dashed curves show the changes with $x,$  for
$m_{\tilde\mu_L}=300\gev.$
\bigskip

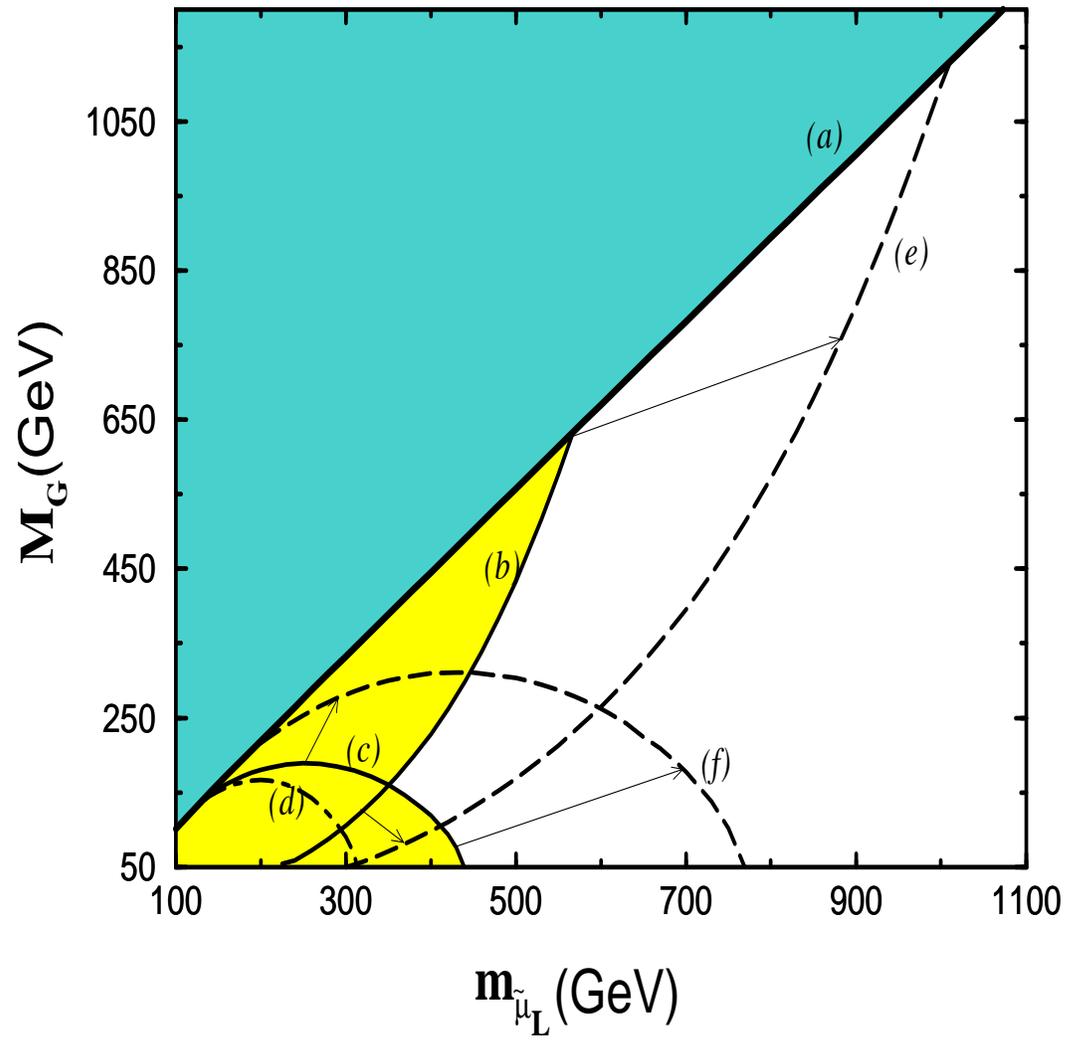
\noindent Figure 7:  $(\mg, m_{\tilde\mu_L})$ parameter space excluded by
present and (possible) future data (all curves are for $\tan\beta=3, \mu>0,
A_0=1).$ Area above line (a) is excluded by the R-G analysis (see discussion in
text). Area between lines (a) and (b) is excluded
by present upper limit on BR(\mueg). Area between line (c) (drawn for
$x=0.5)$ and axes is  excluded by present upper limit on $\taumg.$ Line (d) is
the $x=1$ equivalent to (c). Lines (e) and (f) show the range of parameters
excluded if current limits were decreased by a factor of 10.

\section*{Acknowledgement} This work is supported in part by the National
Research Foundation, under Grant No. PHY-9411546.

\begin{thebibliography}{99}
\bibitem{GJ} H.Georgi and C.Jarlskog, \plett{86}(1979) 297.
\bibitem{FN} C.D. Froggatt and H.B. Nielsen, \np{147}(1979) 277.\  A
recent discussion in the context of SO(10) has been given by G.~Anderson,
S.~Raby, S.~Dimopoulos, L.~J.~Hall, and G.~D.~Starkman, \prd{49}(1994) 3660.
\bibitem{Cha} S. Chaudhuri,\ S. Chung and J. Lykken,\
hep-ph/9405374, FERMILAB-PUB-94/137-T (1994).
\bibitem{Cle}G.~Cleaver, hep-th/9409096, OHSTPY-HEP-T-94-007 (1994).
\bibitem{Moh}K.~S. Babu and R.~N. Mohapatra, \prl{74}(1995) 2418.
\bibitem{HKR} L.~Hall, V.A. Kostelecky and S. Raby, \np{267}(1986) 415.
\bibitem{Mas} F.~Borzumati and A.~Masiero, \prl{57}(1986) 961;\\ F.~Gabbiani
and
A.~Masiero, \np{322}(1989) 235.
\bibitem{Bar} R.~Barbieri, L.~Hall and A.~Strumia, \np{445}(1995) 219.
\bibitem{HK} H.E. Haber and G.L. Kane,{\em  Phys.~Reports} {\bf 117} (1985)
75;\\ H.~P.~Nilles, {\it ibid.} {\bf 110} (1984) 1.
\bibitem{Sug}L.~Hall, J.~Lykken, and S.~Weinberg, \prd{27}(1983) 2359;\\
S.~K.~Soni and H.~A.~Weldon, \plett{126}(1983) 215.
\bibitem{Non} As an example,  Kahler terms of the form
$\int d^4\theta\psi_3^{\dagger}\psi_3(S_2^{\dagger}S_2)/M^2)^n$
in conjunction with
gaugino exchange will  generate a contribution $\delta m^2 \sim m_{1/2}^2.$
\bibitem{Dat}Particle Data Group, {\em Review of Particle Properties.}
\bibitem{AN} R. Arnowitt and P. Nath, \prd{46}(1992) 3981.
\bibitem{MV}M.~E.~Machacek and M.~T.~Vaughn, \np{222}(1983) 831;\ {\bf
B236} (1984) 221;\ {\bf B249} (1985) 70. The importance of working to two loops
has been pointed out in ref. \cite{Kan}
\bibitem{Kan} G.~L.~Kane,\ C.~Kolda,\ L.~Roszkowski,\ J.~D.~Wells,\
\prd{49}(1994) 6173.
\bibitem{Dim2} S.~Dimopoulos, L.~Hall, and S.~Raby, \prd{45}(1992) 4192;\\
G.~W.~Anderson, S.~Raby, S.~Dimopoulos, and L.~Hall, {\it ibid.} {\bf 47}
(1993)
3702.
\bibitem{BB} V.~Barger,\ M.~S.~Berger and P.~Ohmann,\ {\it ibid.}{\bf
D47} (1993) 1092;\\ V.~Barger,\ M.~S.~Berger,\ T.~Han and M.~Zralek, \prl{68}
(1992) 3394.
\bibitem{Neu} M.Neubert,\ preprint CERN-TH/95-107,\ hep-ph/9505238.
\bibitem{Go} S.~G.~Gorishny,\ A.~L.~Kataev,\ S.~A.~Larin and L.~R.~Surgaladze,\
{\em Mod. Phys. Lett} {\bf A5} (1990) 2703.
\eb\ed
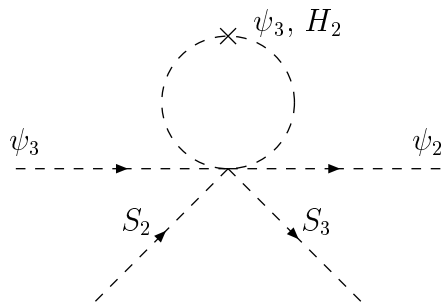
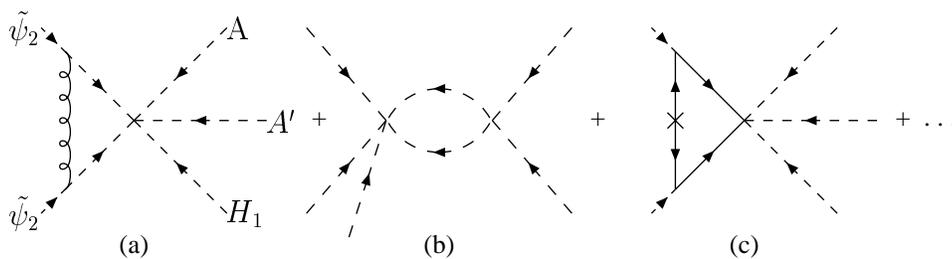
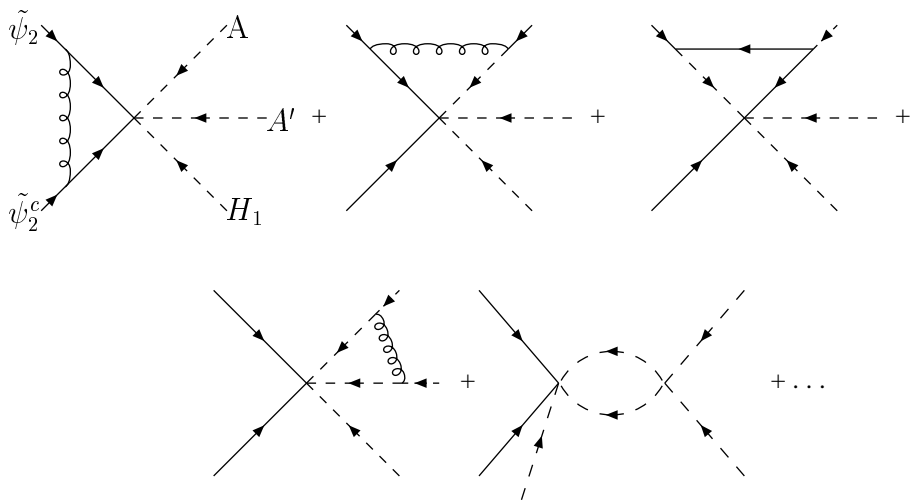
